\documentclass[12pt]{article}
\usepackage{epsfig}
\usepackage{amssymb}
\usepackage{amsmath}
\usepackage{pslatex}
\usepackage[usenames]{color}

\makeatletter

\@addtoreset{equation}{section}
\makeatother
\newcommand{\bea}{\begin{eqnarray}}
\newcommand{\eea}{\end{eqnarray}}
\newcommand{\be}{\begin{eqnarray}}
\newcommand{\ee}{\end{eqnarray}}

\def\tilde{\widetilde}
\def\bar{\overline}

\def\rmd{{\rm d}}
\def\zh{z_{\rm H}}

\begin{document}

\begin{titlepage}
\begin{flushright}
\end{flushright}
\vskip0.2cm
\centerline{\Large \bf  Composite Fermion Metals}
\vskip0.35cm
\centerline{\Large \bf from}
\vskip0.35cm
\centerline{\Large \bf Dyon Black Holes and S-Duality}
\vskip1.5cm
\centerline{\large Dongsu Bak\,,$^a$\, \,\,\,\,
Soo-Jong Rey $^{b,c}$ }
\vspace{1.5cm}
\centerline{\sl $^a$ Department of Physics, University of Seoul, Seoul 130-743 {\rm KOREA}}
\vskip0.25cm
\centerline{\sl $^b$ School of Physics \& Astronomy, Seoul National University, Seoul 141-747 {\rm KOREA}}
\vskip0.25cm
\centerline{\sl $^c$ School of Natural Sciences, Institute for Advanced Study, Princeton NJ 08540 {\rm USA}}
\vskip0.35cm
\centerline{\tt dsbak@uos.ac.kr \,\,\, sjrey@snu.ac.kr}
\vspace{1.5cm}
\centerline{ABSTRACT}
\vspace{0.75cm}
We propose that string theory in the background of quantized dyon black holes in four-dimensional anti-de Sitter spacetime is holographic dual to composite Dirac fermion metal. By utilizing S-duality map, we show that thermodynamic and transport properties of the black hole match with those of composite fermion metal, exhibiting Fermi liquid-like behavior. Built upon Dirac-Schwinger-Zwanziger quantization condition, we propose that turning on magnetic charges to electric black hole along the orbit of $\Gamma_0(2)$ subgroup of $SL(2, \mathbb{Z})$ is equivalent to attaching even unit of {\sl statistical} flux quanta to constituent fermions. Being at metallic point, the statistical magnetic flux is interlocked to the background magnetic field. We find supporting evidences for proposed holographic duality from study of internal energy of black hole and probe bulk fermion motion in black hole background. They show good agreement with ground-state energy of composite fermion metal in Thomas-Fermi approximation and cyclotron motion of constituent or composite fermion excitations near the Fermi-point.

\noindent
\end{titlepage}

\newpage
\section{Introduction}
\vspace{5mm}
The fractional quantum Hall effect refers to an exotic incompressible state of two-dimensional electron system in a high magnetic field, the most well-known occurring at one-third filling, $\nu = 1/3$.
Along with the high temperature superconductivity, the fractional quantum Hall effect has provided a cornerstone of strongly correlated electron systems that deviate from Landau's Fermi liquid behavior. In follow-up developments, a notion was advanced that the fractional quantum Hall effect can be understood in terms of a composite of an electron and even units of magnetic fluxes \cite{CF}. Remarkably, this so-called `composite fermion' \cite{CFreview} turned out not just a theoretical construct but a physical reality. Over the last decade, this proposal has successfully passed a variety of highly nontrivial experimental tests, verifying emergent Fermi sea \cite{emergentFS}, Shubnikov-de Haas oscillations, cyclotron orbits \cite{cyclotron} and magnetocapacitance \cite{magnetocapacitance}. Nowadays, it is even possible to confine a composite fermion inside mesoscopic devices, enabling to observe its semiclassical dynamics \cite{exp}. As such, the composite fermion paradigm is considered a parameter-free, defining theory of strongly correlated electron systems under a high magnetic field.

As the magnetic field is tuned to half-filling, $\nu = 1/2$, a gapless, compressible state emerges out of the fractional quantum Hall system. This state consists of composite fermions, whose fluxes attached to them cancel out the background magnetic field. The composite fermions can propagate freely, thus exhibiting a metallic behavior. Numerous experimental observations of composite fermion in the gapless state around $\nu = 1/2$, where neither a quantum Hall plateau nor an excitation energy gap exist, was a watershed for the concept and physical reality of the composite fermions. Being a highly nontrivial Fermi liquid, various aspects of the `composite fermion metals' were studied in depth. Still, as the flux attachment is intrinsically a nonperturbative process, understanding the composite fermion metal in a theoretically controlled way still poses a challenge. Jain's composite fermion paradigm also predicts companion states around other hierarchical fillings $\nu = 1/2p$ with $p=2,3,\cdots$, at which a constituent electron binds with $2p$ quanta of magnetic flux. They are also gapless and hence compressible. Hereafter, we shall refer all of these hierarchical states as {\sl composite fermion metal}.

The composite fermion metal \cite{KZ} \cite{HLR} is formulated most compactly in terms of so-called fermionic Chern-Simons theory \cite{LF}. This description provided a framework for perturbative approaches and had remarkable successes, but
several major problems have appeared in describing these states as a Fermi liquid. At the outset, these problems are rooted in part to the infrared divergences of Chern-Simons gauge field fluctuations that become particularly acute in the limit of large external magnetic field. Also, various perturbative approaches in the fermionic Chern-Simons theory invoke approximations that often do not ensure self-consistency and maintain gauge invariance. It thus became highly desirable to look for an alternative formulation of the composite fermions in which these difficulties could be successfully overcome and, hopefully, shed new lights.

The purpose of this work is to propose one such alternative formulation of the composite fermion metal. We shall invoke the gauge-gravity correspondence that originated from string theory and argue that hierarchy of the composite fermion metals can be described via holography by the gravity in the background of a hierarchy of dyon black holes -- black holes that carry both electric and magnetic charges -- in (3+1)-dimensional anti-de Sitter spacetime. The assertion is an elementary consequence of our previous observation \cite{bakreyFS} that strongly interacting Dirac Fermi liquid in $d$ dimensional spacetime shares variety of physical properties with electrically charged black hole in $(d+1)$-dimensional anti-de Sitter spacetime. In this holographic duality, the dimension $d=3$ is special in that, on (3+1)-dimensional spacetime, electric and magnetic Maxwell fields are transformable each other by the well-known $SL(2,R)$ S-duality rotation. In the proposed duality, we shall find that semiclassical treatment of electric and magnetic charges in accordance with the Dirac-Schwinger-Zwanziger quantization condition is essential. With the quantization condition, we then identify the relevant duality map for the hierarchy of composite fermion metals, equivalently, of dyon black holes as the $\Gamma_0(2)$ subgroup of $SL(2, \mathbb{Z})$.

We organized this paper as follows. In section 2, we recall the composite fermion paradigm with emphasis on aspects we shall put to holographic dual descriptions. In particular, we focus on composite Dirac fermions, in which fermions are massless. In section 3, we describe solution and properties of dyon black hole in $(3+1)$-dimensional anti-de Sitter spacetime and emphasize that the dyon black hole is related by electric-magnetic duality rotation to a purely electric black hole. In section 4, we invoke the aforementioned charge quantization and analyze duality hierarchy present in the composite fermion metals and dyon black holes. We observe that $\Gamma_0(2)$ subgroup of $SL(2, \mathbb{Z})$ is the relevant duality map of the both. We then propose that dyon black hole is a holographic dual description of the composite fermion metal. In subsequent sections, we put our proposal into tests.
In section 5, we compute the ground-state energy of the composite fermion metal from the dyon black hole description and study its dependence on filling factor $\nu$. We compare it with the result computed earlier from the Hartree-Fock approximation and find an excellent agreement.
In section 6, again from the dyon black hole description, we identify a gauge invariant operator creating a single composite fermion of fermion number and magnetic flux $(N, \Phi)$ in (2+1)-dimensional composite Dirac fermion metal with a charged bulk fermion field carrying both electric and magnetic charges $(q_e,q_m)$ in (3+1)-dimensional background of the dyon black hole. We then study classical dynamics of bulk fermion and show that one-particle dynamics agrees with cyclotron motion of composite fermion.
In section 7, we summarize our results and discuss various issues for further study.

\section{Composite Dirac Fermions}

First we recapitulate salient features of the composite fermion metal \cite{CF, CFreview, KZ, HLR, LF}. We highlight those aspects relevant for the holographic dual in terms of dyon black holes put forward later.

The notion of composite fermion is far more general, but we find it convenient for foregoing discussions to pose it in the context of fractional quantum Hall effect. So, start with a fractional quantum Hall system of constituent fermions of band mass $m_*$ in a constant magnetic
field $B = \nabla \times {\bf A}$ on a planar region of area $L^2$.
The composite fermions are made by attaching fluxes to fermions in the uppermost Landau level while the lower Landau levels are treated as an inert incompressible background. Ignoring interactions between the constituent fermions,
the one-particle spectrum breaks up into evenly spaced Landau levels with energy
\bea
E_\ell = \hbar \omega_c (\ell + {1 \over 2}) \qquad \mbox{where} \qquad
\omega_c = {e\,\, B \over m_* c}.
\eea
For a fixed band mass $m_*$,
the cyclotron frequency $\omega_c$ increases linearly with $B$.
Each Landau level has  a degeneracy $(B/\Phi_0)$ where
$\Phi_0 = {2 \pi \hbar c / e}$
is the magnetic flux quantum. The filling fraction, defined as
\bea
\nu := {\Phi_0 \over B} \ \rho
\eea
where $\rho := N/L^2$ is the constituent fermion number density,
indicates that $[\nu]$ is the number of
Landau levels completely filled.
When $\nu$ is an integer, there is a
discontinuity in the chemical potential, leading to an incompressible integer quantum Hall state. When $\nu$ is a fractional value, because of the degeneracy of one-particle states, the physics is controlled by interaction energy measured at the magnetic length
$
\ell_B = \sqrt{ {\Phi_0 / 2 \pi B}}$
.
Of particular interest is a large magnetic field regime, where $\nu < 1$. In this case, the interaction energy scale is much less than the cyclotron scale. For rational values of $\nu$ with odd-denominator, the
system becomes an incompressible fractional quantum Hall state. For rational values of $\nu$ with even-denominator, the system remains compressible and behaves as a Fermi liquid, with distinguishing features that were verified experimentally. In both cases, because of the large degeneracy of states, standard perturbation theory with respect to the interaction potential becomes ineffective. It is this
nonperturbative nature that renders the physics of both cases extremely rich and interesting.

To compare with dyon black hole, we consider a (2+1)-dimensional system with underlying Lorentz invariance. In the black hole, dyon gauge field background breaks the SO(3,2) isometry of the background anti-de Sitter spacetime explicitly to ISO(2) Euclidean and U(1) gauge symmetry. This means that corresponding (2+1)-dimensional system must have a homogeneous distribution of U(1) charge density. The underlying SO(3,2) isometry is much larger than Galilean or Schr\"odinger symmetries of conventional quantized Hall systems such as high mobility GaAs semiconductors. This means that for comparison with dyon black hole we need to substitute the band mass $2 m_*$ by the energy $E/c^2$ and obtain the relativistic
Landau levels with Fermi velocity $v_{\rm F} = c$ and energy
\bea
E_\ell = \Omega_c \sqrt{2\ell} \qquad \mbox{where} \qquad \Omega_c = \sqrt{c \hbar e B}.
\eea
Notice that the characteristic
frequency $\Omega_c$ is proportional to {\sl square-root} of the magnetic field $B$. On the other hand, the filling-fraction $\nu$ and hence the magnetic length $\ell_B$ remain the same as the Galilean invariant
case. So, at $\nu$ with even denominator, the
system is compressible and behave as a relativistic Fermi
liquid, which we shall call Dirac Fermi liquid. We thus have a composite Dirac fermion metal at a quantum critical point. Note that the Landau Fermi liquid itself can be considered as a system at criticality in the sense that all correlations decay algebraically. It suggests that the composite fermion metal may be treated as a critical, relativistic Landau Fermi liquid.

The flux attachment can also be described by a U(1) Chern-Simons theory of statistical gauge field coupled to the constituent fermions. In this case, the composite fermions are specifically named as Chern-Simons fermions.
In the Dirac system under consideration, we take the constituent fermions as two-component massless
Dirac fermions and consider the Chern-Simons Lagrangian:
\bea
L_{\rm CFT} &=& \int \rmd^2 {\bf r} \
\Big[ \overline{\psi} \,\,\gamma^m ( i \partial_m + {e \over c} A_m -
a_m) \psi +
{\nu \over 4\pi} \epsilon_{mnp} a_m \partial_n a_p \Big]
\nonumber \\
&+& {v_{\rm F}
 \over 4 \pi}
\int \rmd^2
{\bf r} \int
\rmd^2 {\bf r}' \ \overline{\psi} ({\bf r}) \psi({\bf r})
{g \over |{\bf r} - {\bf r}'|} \overline{\psi} ({\bf r}') \psi({\bf r}').
\eea
Here, ${\bf A} := (0, - y/2, + x/2) B$ is the background magnetic field, $(a_0, {\bf a})$ is the {\sl statistical} magnetic potential for the flux attached to the fermions, $\nu/4\pi$ is the Chern-Simons coefficient determining units of the flux attached, and $g$ measures the scale-invariant density-density interaction. For the fermion,
$\gamma^m$ are Dirac gamma matrices, $\overline{\psi}= \psi^\dagger \gamma^0$ is the charge conjugate and $v_{\rm F}$ is the Fermi velocity.
Notice also that the dispersion relation of the constituent fermion is linear, $\omega = v_{\rm F} |{\bf k}|$, where $v_{\rm F}$ denotes (renormalization group fixed-point value of) the Fermi velocity.

The fermion - statistical flux relation
follows from the Lagrange constraint equation of $a_0$
\bea
\rho ({\bf r}, t) =
{\nu }\, \, {b({\bf r}, t)\over e\,\Phi_0} \ ,
\label{constraint}
\eea
where
\bea
b({\bf r}, t) = \langle \nabla \times {\bf a} \,\rangle
\eea
and
\bea
\rho ({\bf r}, t) :=
 \Big\langle \psi^\dagger ({\bf r}, t) \psi({\bf r}, t) \Big\rangle =
\sum_{a=1}^N  \ \delta ({\bf r}- {\bf r}_a (t)).
\eea
For the composite fermion with $\Phi = 2p \Phi_0$ flux quanta  attached,
one is setting $\nu$ to $\nu_c = 1/2p$. If the statistical gauge potential ${\bf a}$ is chosen as
\bea
{\bf a} ({\bf r}, t) = {1 \over \nu_c} 
\sum_{a=1}^N {\hat{\bf z} \times ({\bf r} -
{\bf r}_a(t)) \over |{\bf r} - {\bf r}_a(t)|^2}\ ,
\eea
we see that the constraint (\ref{constraint}) is satisfied. As in Galilean invariant situation, the wave function $\Psi$ of the Chern-Simons Dirac fermions obeying the above constraint equation are related to the wave function
$\Psi_0$ of the constituent Dirac fermions as
\bea
\Psi (z_1, \cdots, z_N) = \prod_{a<b} \left( {(z_a - z_b) \over |z_a - z_b|} \right)^{1 \over \nu_c} \
\Psi_0 (z_1, \cdots, z_N) \ ,
\eea
where $z_a = x_a + i y_a$ is the position of $a$-th fermion.

The simplest approach to study the system is the mean field approximation: the fermion number density is assumed spatially homogeneous. By the constraint (\ref{constraint}), the Chern-Simons flux
quanta attached to the constituent fermions are also smeared out to a homogeneous statistical magnetic field. As the
Chern-Simons flux was arranged in the opposite orientation to the external magnetic field $B$,
the effective magnetic field $\Delta B$ experienced by the constituent fermion is given by
\bea
\Delta B:= B - {1\over e}
\langle \nabla \times {\bf a} \rangle = B - {\Phi_0 \over \nu_c} \ \rho.
\label{deltaB}
\eea
We see that $\Delta B$
cancels out identically to zero with the Chern-Simons filling fraction
\bea
\nu := {\Phi_0 \over B } \ \rho = \nu_c:= {1 \over 2p}.
\label{metalrelation}
\eea
At this special filling fraction, the Chern-Simons theory is describable in the mean-field approximation
in terms of Dirac fermions at zero magnetic field. The system should be a compressible Fermi-liquid like state. For Galilean invariant fermions, the existence of Fermi-liquid like state at $\nu = \nu_c$ was predicted by Kalmeyer and Zhang \cite{KZ} and by Halperin, Lee and Read \cite{HLR}. It is now clear why the composite fermions or Chern-Simons fermions is the right starting point. The Fermi-liquid like state of composite fermions is a nondegenerate state. Unlike the original highly degenerate Landau level state, the composite fermion state provides a nondegenerate starting point of well-posed perturbation theory. That this is the right starting point was ultimately justified by agreement with numerous experimental results and exact diagonalization studies.

If the filling fraction is away from $\nu_c$, the external magnetic field and the Chern-Simons flux do
not cancel out. In the mean field description, the system is described as non-interacting fermions in the
residual, effective magnetic field $\Delta B$. The effective filling fraction for these composite fermions
is given by
\bea
n = {\Phi_0 \over \Delta B} \ \rho
\eea
When $n$ is integer-valued, at the mean field level, this is just the system of $n$ filled Landau levels of composite fermions. This observation is the basis of the composite fermion paradigm: the fractional quantum Hall effect of constituent fermions is viewed as the integer quantum Hall effect of composite fermions.
In fact, from (\ref{deltaB}), we see that the filling fraction $\nu$ of the constituent fermions is given by
\bea
\nu = {n \over {n / \nu_c} + 1} \qquad \mbox{equivalently} \qquad
{1 \over \nu} - {1 \over \nu_c} = {1 \over n}.
\label{qhrelation}
\eea
The excitation energy gap for these quantum Hall states are naturally given by the corresponding effective characteristic
 frequency of the composite fermions
\bea
\Delta E = \hbar \Delta \Omega_c = \sqrt{c \hbar e \Delta B}.
\eea

For further consideration in comparing with the dyon black hole, we stress by now the obvious point that the composite fermions are at a metallic point when the total statistical gauge flux and total background magnetic flux are equal. See (\ref{metalrelation}). Therefore, at least at the level of mean-field approximation, the two gauge fields and hence their magnetic flux quantum numbers are interlocked each other. This is significant to our foregoing considerations has the following immediate implication: in a holographic relation of the composite fermion to the dyon black hole, a metallic point is uniquely specified by {\sl one} quantum number whereas a gapped quantum Hall point is specified by {\sl two} quantum numbers. This is manifest in (\ref{qhrelation}).

\section{Dyon Black Holes}

In this section, we turn to dyon black hole in four-dimensional anti-de Sitter spacetime. The dyon black
hole is the most general classical solution carrying electric charge 
and magnetic charge 
in
the Einstein-Maxwell system
\bea
I = \gamma \int \rmd^4 x \sqrt{-g} \ \Big( R_{(4)}
+ {6 \over \ell^2} - {1 \over e_G^2} g^{ab} g^{cd} F_{ac} F_{bd}  -
\theta {1 \over \sqrt{-g}} \epsilon^{abcd} F_{ac} F_{bd} \Big) \ .
\label{action}
\eea
Here, $\gamma = 1/(16 \pi G_4)$ is the inverse of Newton's constant, $e_G, \theta$ are dimensionless Maxwell coupling constant and $P, T$-violating theta angle, respectively. Below we shall set $e_G=1$, $\theta=0$
and, instead, introduce a suitable conversion factor (which will be denoted as $\sigma_F$) and boundary conditions later on.

The dyon black hole is described by the spacetime metric
\bea
\rmd s^2 &=& {\ell^2
\over z^2} \Big[ -f(z) \rmd t^2 + {1 \over f(z)} \rmd z^2 + \rmd x^2 + \rmd y^2 \Big] \nonumber \\
&& f(z) = 1 - {1 \over 2} m \ z^3 + {1 \over 16} (\rho_e^2 + \rho_m^2) \ z^4
\label{metric}
\eea
and the U(1) Maxwell field strength
\bea
F = {1 \over 4} \Big( \rho_e \,\rmd t \wedge \rmd z +
\rho_m \,\rmd x \wedge \rmd y \Big) \ .
\label{F}
\eea
For some other holographic applications of dyon black holes to condensed matter systems, see \cite{dyonBH}. For other holographic approaches dealing with quantum Hall effect and plateau transitions, see \cite{holoQHE}.

The black hole is characterized by four parameters $\ell, m, \rho_e,
\rho_m$. The parameter $\ell$ sets the curvature scale of the
anti-de Sitter spacetime, $m$ sets the total mass (energy) density, and $(\rho_e, \rho_m)$ denotes the electric and magnetic number densities, respectively.

The black hole is planar along $x,y$ directions and is centered at $1/z = 0$, where the spacetime metric is singular. The singularity is shielded by the black hole horizon
$z=\zh$, which is the location of the outermost zero of the metric function $f(z)$:
\bea
1 - {1 \over 2} m \zh^3 + {1 \over 16} (\rho_e^2 + \rho_m^2) \zh^4 = 0.
\eea
Such a zero is guaranteed to exist
at finite $z$ provided $m, \rho_e,\rho_m$ obey the inequality
\bea
\Big({1 \over 3} \big(\rho_e^2 + \rho_m^2\,\big) \Big)^3 \,\, \le
\  m^4\ .
\label{inequality}
\eea
In gravity, this requirement is referred as the {\sl cosmic censorship}: the singularity must be shielded by an event horizon.
In terms of $(\zh, \rho_e, \rho_m)$, we can express the
metric function (\ref{metric}) in a convenient form as
\bea
f(z) = 1 - \Big({z \over \zh}\Big)^3 +
{1 \over 16} (\rho_e^2 + \rho_m^2) \zh^4 \ \left[ \Big({z \over \zh}\Big)^4 - \Big({z \over \zh}\Big)^3 \right] \ .
\eea

Salient features of the dyon black hole that match with features of the composite Dirac fermion metal are
the followings.
\begin{list}{$\bullet$}{}
\item Though metric (\ref{metric}) exhibits underlying SO(3,2) asymptotic isometry, the U(1) gauge field (\ref{F}) breaks this to E(2) Euclidean isometry. The asymptotic U(1) gauge symmetry is interpretable as holographic realization of conserved fermion number of the composite Dirac fermion metal. This means that a holographic dual must be a relativistic system, where a {\sl homogeneous} density and temperature of the metal.


\item The dyon black hole carries both electric and magnetic charges, so it breaks parity $P$ and time-reversal $T$ in $(3+1)$ dimensional anti-de Sitter spacetime. This also means that, in a $(2+1)$-dimensional holographic dual system defined at asymptotic infinity, $P, T$ are broken. Nevertheless, the dyon black hole is static. This property is holographically dual to the
    property that composite fermion system is non-rotating though fermions carry nonzero fluxes.

    Normally, in $P, T$-violating situations by electromagnetic fields, a nonzero value of field
    angular momentum
\bea
{\bf L} = \gamma \int_{\rm AdS_4} \rmd^3 {\bf r} \ {\bf r} \times ({\bf E} \times {\bf B})
\eea
would be present as well. For dyon black holes under consideration, we see that the field angular momentum ${\bf L}$ vanishes identically. This is very different from the configurations obtained by bringing an electric charge to a magnetically charged black hole (or vice versa). In the latter, at nonzero separation between the electric charge and the magnetic black hole, the field angular momentum is nonzero. It can be shown \cite{garfinklerey} that, after the electric charge falls inside the magnetic black hole, the field angular momentum is transmuted to an intrinsic spin of the dyon black hole. In other words, the final configuration is spinning and hence cannot be spherically symmetric. By black hole no hair theorem, the final configuration ought to be a Kerr-Newman black hole in anti-de Sitter spacetime. This is different from ours.

\item The dyon black hole carries magnetic charge as well as electric charge. The magnetic charge is interpretable as holographic realization of the statistical gauge flux attached to constituent fermions in forming composite fermions. First of all, as stress in the previous section, total statistical gauge flux is interlocked to the total external magnetic flux. Moreover, field equations of the Einstein-Maxwell system are invariant under the S-duality that continuously rotates
the Maxwell electric and magnetic field strengths each other:
\bea
{\bf B} \ \rightarrow \ \widetilde{\bf B} = \sin \alpha \ {\bf B} - \cos \alpha \ {\bf E} \nonumber \\
{\bf E} \ \rightarrow \ \widetilde{\bf E} = \cos \alpha \ {\bf B} + \sin \alpha \ {\bf E}
\label{sduality}
\eea
Similarly, the S-duality rotates the electric and magnetic charge densities $(\rho_e, \rho_m)$ to $(\tilde{\rho}_e, \tilde{\rho}_m)$. The duality rotation is holographically dual to the duality between
fermions and fluxes well known in Jain's construction of the composite fermions.

We also see that a dyon black hole can be rotated to a purely electric black hole, which takes exactly the same form as (\ref{metric}) - (\ref{hawkingtemperature}) except that the charge densities $(\rho_e, \rho_m)$ are replaced by rotated ones, $(\widetilde{\rho}_e, \widetilde{\rho}_m) = (\sqrt{\rho_e^2 + \rho_m^2}, 0)$. We see likewise that thermodynamics of dyon black hole is covariant under the
S-duality transformation.

\item Thermodynamics of the dyon black hole matches well with thermodynamics of the composite Dirac fermion metal. To discuss thermodynamics, we take the dyon black hole in a large box of size $L^2$ ($L \gg \ell)$.
The total energy (mass) is given by
\bea
E =  m \gamma\,\, L^2 \ ,
\eea
the total entropy is given by
\bea
S = {4\pi \over \zh^2} \gamma\,\, L^2 \ ,
\eea
and the total electric and magnetic charges are given by
\bea
Q'_e = \rho_e \gamma\,\, L^2 \qquad \mbox{and} \qquad Q'_m =
\rho_m \gamma\,\, L^2  \ .
\label{charges}
\eea
The dyon black hole is in equilibrium at Hawking temperature
\bea
T_{\rm H} = \ {1 \over 4 \pi\,\zh} \
\Big(3  - {\zh^4 \over 16} (\,\rho_e^2 + \rho_m^2) \Big) \ .
\label{hawkingtemperature}
\eea
In the canonical ensemble where an appropriate combination of the charge densities $(\rho_e, \rho_m)$ is fixed, chemical potential $\mu$ of the dyon black hole is determined in terms of the electric charge density as
\bea
\mu := A_0 (z=0) = {1 \over 4} \rho_e \zh \ .
\eea

It is well known that black holes obey laws of thermodynamics. Previously, from careful study of thermodynamic and transport properties, we found that purely electric black hole ($\rho_m=0$) in $(d+1)$-dimensional anti-de Sitter spacetime behaves very similar to ideal Dirac Fermi liquid in $d$-dimensional flat spacetime. There, we interpreted Coulomb repulsion of constituent electric charges in black hole as a direct holographic manifestation of the Pauli exclusion principle among constituent fermions in the Fermi liquid. In Fermi liquid, the {\sl quantum} Pauli pressure increases the ground-state energy. In black hole, the {\sl classical} Coulomb repulsion increases black hole binding energy. This is the origin of the cosmic censorship condition, which is given for electric black hole by (\ref{inequality}) with $\rho_m=0$. Built upon this observation, we proposed that charged black holes in anti-de Sitter spacetime is holographically dual to fermi Dirac liquid.

For the dyon black hole under consideration, we again find that many features of its thermodynamics are similar to those of ideal Fermi liquid. This roughly fits to the emergent Fermi surface of composite fermions already observed experimentally \cite{emergentFS}. The cosmic censorship condition (\ref{inequality}) implies that minimum of the energy density $m$ is set by the electric and magnetic charge densities. The condition is strictly above the bound at finite temperature and becomes saturated at zero temperature.
 So, at zero temperature,
 the energy density and the charge density are related each other. Translated to quantities in composite
 fermion metal, it relates the energy density
$\varepsilon'$ to fermion number density $\rho'$ (in gravity units)
as well as
 the flux quanta per fermion number $\Phi/N$:
\bea
\varepsilon' = \Big( {1 \over 3} (\rho')^2 \
\Big[1 + {\Phi^2 \over N^2} \Big] \Big)^{3 / 4}.
\eea
The exponent on the right-hand side is precisely the scaling behavior of the ideal fermi Dirac liquid in $(2+1)$-dimensional spacetime {\sl provided} $\Phi/N$ is fixed, equivalently, $\rho_e$ and $\rho_m$ are both related to the fermion number density. This must be so since the right-hand side of (\ref{inequality}) was interpreted as holographic manifestation of the Pauli pressure and the Pauli exclusion principle arises from fermion statistics. In the next two sections, we shall substantiate this observation further, both at zero and finite temperature and identify composite fermion metal as the system behaving as an ideal fermi Dirac liquid, where $\rho_e$ is mapped to the fermion number density while $\rho_m$ is mapped to the flux density attached to each fermion.

\end{list}

\section{Holography of S-duality}

We mentioned already that the dyon black holes can be transformed one another by S-duality rotations. In this section, we shall study further aspects of the S-duality and argue that, by treating electric and magnetic charges of the dyon black hole quantum mechanically, the $SL(2, \mathbb{Z})$ S-duality map of the black hole coincides with the $SL(2, \mathbb{Z})$ flux-charge duality map of the composite Dirac fermions. We
shall then identify $\Gamma_0(2)$ as the relevant subgroup of $SL(2, \mathbb{Z})$ for the composite
fermion metals. The modular group $SL(2, \mathbb{Z})$ was previously related to the transformations among various quantum Hall plateaus \cite{dolan}.
The S-duality was studied previously in the context of AdS/CFT correspondence \cite{Witten}. In the present context, we believe our presentation below is much simpler and more intuitive. We closely follow \cite{shaperewilczek} and, in particular, \cite{reyzee}.

\subsection{composite Dirac fermions}
We shall now argue that, in the composite Dirac fermion description, fermion's charge and flux can be mapped to each other. We do so by recasting how the statistical gauge field $a_m$ arose for the Chern-Simons description in section 2. Start with Dirac fermions, possibly interacting with an external electromagnetic gauge potential $A_m$ that breaks $(2+1)$-dimensional parity $P$ and time reversal $T$. The fermion number is a conserved quantity and the corresponding current density $j^m$ obeys $\partial_m j^m = 0$. Having the fermions living in (2+1) dimensions, we can express the current vector $j^m$ in terms of a newly
introduced dual vector $a_m$
\bea
j^m = {1 \over 2 \pi} \epsilon^{mnp} \partial_n a_p \ .
\eea
The relation between $j^m$ and $a_m$ is not unique. We can transform $a_p$ inhomogeneously to a new dual vector $a_p - \partial_p \Lambda$ for an arbitrary scalar $\Lambda$ and still describes  the same
$j^m$. We see that any description of the Dirac fermions in terms of dual vector $a_m$ must be gauge invariant as well as conformally invariant.
Recalling that the composite fermions live in an external magnetic field created by the background gauge potential $A_m$ and hence break $P$ and $T$,
this leads to an equivalent theory
\bea
L &=& \int \rmd^2 {\bf r} \ \Big[ \ {1 \over 2} \Big({\nu_{\rm D} \over 2 \pi} \varepsilon^{mnp} a_m \partial_n a_p + {1 \over \lambda_{\rm D}} f_{mn} {1 \over \sqrt {-\partial^2}} f_{mn} \Big) + a_m \tilde{j}^m + A_m j^m
\ \Big] \ ,
\label{equivalent}
\eea
labeled by the Chern-Simons coefficient $\nu_{\rm D}$ and the coupling parameter $\lambda_{\rm D}$. Here, $j^m$ denotes conserved current for the Dirac fermions and $\tilde{j}^m$ denotes conserved current for fluxes. Accordingly, $A_m$ is the gauge potential acting as a source for the fermion current $j^m$ and $a_m$ is the statistical gauge potential acting as a source for the flux current $\tilde{j}^m$. In the equivalent theory (\ref{equivalent}), we included all possible gauge invariant and conformal invariant terms of the dual vector field $a_m$. Notice that the two quadratic terms of the dual vector field $a_m$ are $P, T$ even and odd terms, respectively. Even if not present at classical level, both terms are generated once quantum fluctuations are taken into account.

We might have opted to describe the current $\tilde{j}^m$ in terms of another dual vector $b_m$
\bea
\tilde{j}^m = {1 \over 2 \pi} \epsilon^{mnp} \partial_n b_p
\eea
and treat the current $j^m$ as an external source. We then obtain an equivalent theory which takes
exactly the same as (\ref{quadratic}) except the statistical gauge potential $a_m$ is replaced by $b_m$
and the coupling parameters take different values. This description interchanges the role and interpretation of $j^m$ and $\tilde{j}^m$, viz. interchanges the charges and the fluxes of composite fermions.

It is now easy to see that the charge and the flux are dual each other.
We first integrate out the statistical gauge potential $a_m$. In covariant gauge, we get
\bea
L = \int \rmd^2 {\bf r} \ \Big[ \ {1 \over 2} J^{\ m} \ \Big( {{\nu} \over 2 \pi}{\epsilon_{mnp} \partial^p \over \partial^2} + {1 \over {\lambda}} {\delta_{mn} \over \sqrt{-\partial^2}} \Big) \ J^{\ n} \ \Big]
\label{quadratic}
\eea
where
\bea
J^{\ m} = \tilde{j}^m + {1 \over 2 \pi} \epsilon^{mnp} \partial_n A_p
\eea
and
\bea
\Big({\nu}_{\rm D} + i {2 \pi \over {\lambda}_{\rm D}}\Big) \quad \rightarrow \quad \Big( \nu + i {2 \pi \over \lambda} \Big) = - \Big( {\nu}_{\rm D} + i {2 \pi \over {\lambda}_{\rm D}}\Big)^{-1}.
\label{S}
\eea
Expanding the current-current interactions, we find that
\bea
\langle \tilde{j}_m (x) \tilde{j}_n (y) \rangle =
\langle x \vert \Big[ \ {1 \over \lambda_{\rm D}} (\delta_{mn} \partial^2 - \partial_m \partial_n){1 \over \sqrt{-\partial^2}}
+ {\nu_{\rm D} \over 2 \pi} \epsilon_{mnp} \partial_p \ \Big] \vert y \rangle
\eea
is specified by the coupling parameters $(\nu_{\rm D} / 2 \pi + i / \lambda_{\rm D})$, while
(using $\epsilon_{mnp}$-tensor identities repeatedly)
\bea
\langle j_m (x) j_n (y) \rangle
= \langle x \vert \Big[ \ {1 \over \lambda} (\delta_{mn} \partial^2 - \partial_m \partial_n){1 \over \sqrt{-\partial^2}}
+ {\nu \over 2 \pi} \epsilon_{mnp} \partial_p \ \Big] \vert y \rangle.
\eea
is specified by the coupling parameters $(\nu/ 2 \pi + i / \lambda)$, respectively. That is, the
correlation functions of charges and fluxes are dual each other in the sense that the coupling parameters are inverted by (\ref{S}).

In the above considerations, the equivalent theory (\ref{equivalent}) is not unique but comes with an
infinite sequence of an even-integer shift of the Chern-Simons coefficient
\bea
\Gamma_0(2): \qquad \Big( {\nu}_{\rm D}+ i {2 \pi \over {\lambda}_{\rm D}}\Big)
\quad \rightarrow \quad
\Big( {\nu} + i {2 \pi \over {\lambda}}\Big) = \Big( {\nu}_{\rm D} + 2\mathbb{Z} + i {2 \pi \over {\lambda}_{\rm D}}\Big) \ .
\label{T}
\eea
This is interpretable as another type of duality in that one {\sl fermionic} theory is mapped to another self-similar but distinct {\sl fermionic} theory where a different amount of the statistical gauge flux is attached to each fermion. Here, to keep the statistics the same, we only considered the shift by even integral units.

We see that the two transformations (\ref{S}, \ref{T}) form a closed group of duality map. In fact, it is the
$\Gamma_0(2)$ subgroup of $SL(2, \mathbb{Z})$, where the latter is the group generated by (\ref{S}) and
(\ref{T}) permitting shifts by all integral units.

\subsection{dyon black hole}
In the previous section, the dyon black holes were considered as a classical configuration. In particular, the electric and magnetic charges $Q_e, Q_m$ in (\ref{charges}) were treated as {\sl continuous} and {\sl independent} quantities. As well-known, such treatment breaks down at quantum level. The electric and magnetic charges $Q_e, Q_m$, as measured in unit of $e_G$ and $1/e_G$ (both of which we set to unity), are noncommuting
and subject to the Dirac-Schwinger-Zwanziger quantization condition
\bea
Q_e Q_m =
\mathbb{Z}
\label{quantization}
\eea
in our unit system. As it stand, the quantization allows the ratio $Q_e/Q_m$ to be an arbitrary rational number. In this section, we shall present arguments advocating the viewpoint that only particular fraction $Q_e/Q_m =1/2p$ is the relevant value for the composite fermion metal.

We now examine implications and consequences of the quantization condition (\ref{quantization}) to the electric and magnetic charges $(Q_e, Q_m)$ of the dyon black holes. Since the black hole is planar symmetric, we might as well deal with $(\rho_e, \rho_m)$, which are related to $(Q_e, Q_m)$ by the conversion factor $\gamma L^2$. See (\ref{charges}).

Quantum mechanically, the quantization condition breaks the $SL(2, \mathbb{R})$ invariance of the classical Einstein-Maxwell system to $SL(2, \mathbb{Z})$. They are generated by inversion $S$ and shift $T$:
\bea
&& S: \hskip1.2cm Q_e \rightarrow \ Q_m \hskip2.15cm \mbox{and} \qquad  Q_m \rightarrow - Q_e \nonumber \\
&& T: \hskip1.2cm Q_e \rightarrow \ Q_e + p \ Q_m \qquad \mbox{and} \qquad Q_m \rightarrow + Q_m ,
\eea
where $p$ is an arbitrary integer.
These transformations map a point in the two-dimensional charge lattice $(Q_e, Q_m)$ to another point. We see that $T^2$-transformations map even / odd charge lattice sites to itself, thus form a closed subgroup. The subgroup of $SL(2, \mathbb{Z})$ generated by $S$ and $T^2$ is the $\Gamma_0(2)$.

The $S$ transformation is a direct consequence of the S-duality rotation (\ref{sduality}) at $\alpha = 0$. This also means that we are holding ${\bf E}$ and $\widetilde{\bf E}$ as the canonical momentum of the gauge potential,
respectively, before and after the
duality rotation. The $T$ transformation is of different nature. In finding the dyon black hole solution, we started with the $P, T$ violating $\theta$ angle in (\ref{action}) to zero. Under $T$ transformation, the
canonical momentum of the gauge potential changes from ${\bf E}$ to ${\bf E} + p {\bf B}$.

Among the S-duality orbits, we focus on the dyon black holes generated by acting $S T^{2p} S$ on $(Q_e, Q_m)$ black hole for $p=1, 2, \cdots$. We see all of them can be generated by the $\Gamma_0(2)$ subgroup.
Start with electrically charged black hole of charge $(-Q_e, 0)$. The $S$ transformation maps this to $(0, Q_e)$. Further transforming with $T^{-2p}$, the black hole is mapped to $(2p \ Q_e, - Q_e)$. After $S$ transformation, we obtain the black hole of charge $(Q_e, 2p \ Q_e)$.

Can it be that the S-duality group is just the $\Gamma_0(2)$, not $SL(2, \mathbb{Z})$? Here we will make only a speculative remark. Though the field equations are covariant under $SL(2, \mathbb{Z})$, there is a subtlety having to do with global structure of the four-dimensional spacetime. It is known that Euclidean partition function is invariant under $T$ if the four-manifold $X_4$ is a spin manifold, whereas it is invariant under $T^2$ only if the four-manifold $X_4$ is not a spin-manifold. The latter option is certainly viable {\sl if} holographic dual of composite fermion metals do not contain any neutral fermion fields (which describes gauge-invariant fermionic operators in the composite fermion metals). Indeed, in Jain's
construction, the building blocks are electron, hole and flux creation operators. They carry electron number $-1, +1, 0$, respectively, so there is no gauge-invariant composite operators
carrying  fermion number. Moreover, we are attaching only {\sl even} numbers of flux quanta to electrons, so the composites are again fermions -- hence, entire hierarchy of the composite fermions proceeds the same way.


\subsection{holographic duality}
We see that the $(2+1)$-dimensional composite Dirac fermion metals and the $(3+1)$-dimensional anti-de Sitter dyon black holes share the $\Gamma_0(2)$ as the natural S-duality map among hierarchical states. In fact, the relations can be made more formal by carefully analyzing boundary conditions of the gauge potential at asymptotic infinity, as was detailed in \cite{Witten}. For the present discussions, it suffices to recall that, according to the AdS/CFT correspondences, the action $I(\Phi_0)$ of a bulk field $\Phi$ (which includes not only the metric $g_{mn}$ and the gauge potential $A_m$ but also extra probe fields introduced by hand) with boundary condition $\Phi(z=0) = \Phi_0$ is precisely the holographic dual to the equivalent theory we discussed in (\ref{equivalent}) and (\ref{quadratic}).
There, for instance,  a
$T$ transformation
in the bulk side induces a shift of $\nu_D$ to $\nu_D+1$ in the boundary
side.
Thus, our proposal is that the $\Gamma_0(2)$ duality map of the composite fermion metal and the $\Gamma_0(2)$ duality map of the dyon black hole are in fact one and the same. In this sense, the dyon black hole is a new effective description for the equivalent theories of composite Dirac fermion metals.

We can check immediately that our proposal passes several consistency checks and also bears interesting implications.
\begin{list}{$\bullet$}{}

\item Symmetries of the dyon black hole and the composite fermion metal do match. As discussed in the previous section, the dyon black hole breaks the $SO(3,2)$ isometry of the $(3+1)$-dimensional anti-de Sitter spacetime explicitly to E(2) Euclidean symmetry and U(1) gauge symmetry. The $(2+1)$-dimensional composite Dirac fermion metal is gapless and compressible and has the same symmetry breaking pattern. The U(1) electromagnetic field background is dual to a homogeneous, finite density of the composite fermions. The dyon black hole, with its electric and magnetic charges, breaks $P$ and $T$, much the same way as the composite fermions do.

\item As mentioned above, in \cite{bakreyFS}, we proposed that $(\rho_e, 0)$ black hole is holographically dual to a strongly correlated Dirac Fermi liquid whose total fermion number density equals to $\rho_e$
and total energy density equals to $m$. In particular, we showed that thermodynamics of the black hole scales exactly the same as thermodynamics of ideal Fermi liquid. Though such scaling seems peculiar for a strongly correlated system, we pointed out that it fits precisely with the hidden Fermi liquid (HFL) paradigm put forward in the context of high temperature cuprates and of composite fermions \cite{HFL}. In the next section, we shall use this scaling relation to extract ground-state energy of the composite fermion metal and compare it with the Hartree-Fock approximation. The scaling relation of thermodynamics holds universally for all spacetime dimensions, but a special respect takes place in the $(3+1)$ dimensions we are presently interested in. We already explained that all dyon black holes belong to the $SL(2, \mathbb{Z})$ orbits. So,
we see immediately that thermodynamics of dyon black holes must scale the same way as the electrically charged black holes once charges and gauge field strengths are transformed covariantly by the S-duality transformations.


\end{list}

\section{Holography of Ground-State Energy}
For non-relativistic composite fermion system, the ground-state energy was computed in the Hartree-Fock approximation. This is the approximation scheme valid at high density regime. In this regime, the fluxes are treatable as a uniform background and the constituent fermions live in this background. In the Hartree-Fock approximation, the ground-state energy was found \cite{HF} to scale with the
filling factor as
\bea
E_{\rm HF} = E_{0} \Big( 1 + 3 p^2 \Big) \qquad \mbox{where} \qquad
E_{\rm 0}  = {N \over 2} E_{\rm F}.
\label{HFgs}
\eea
This shows that the ground-state of the system exhibits the Fermi surface, fitting with the experimental observation \cite{emergentFS}. Note that this result agrees with the Hartree-Fock ground-state energy of composite fermions at filling fraction $\nu = n/(2p n + 1)$ in the limit $n \rightarrow \infty$, viz. the number of occupied Landau levels in the effective field goes to infinity.

For the ground state of our relativistic system, $E_0$ in the above will be replaced by
\be
E_0 = {2\over 3}\, N\, k_{\rm F}  \qquad \mbox{where} \qquad k_{\rm F} = \sqrt{2 \pi \rho} \ ,
\ee
while the scaling of the leading Hartree-Fock correction remains unchanged.

Let us now turn to gravity side to check the scaling of the
above correction. The total energy of the black hole is given by
\be
E=  m\, \gamma \ L^2\ .
\ee
In order to match the energy $E$ of the gravity side with that of
the weakly coupled conformal fermion as in Ref.~\cite{bakreyFS},
we fix $G_4$ by $1/16\pi G_4 = c_{\rm eff}/6\pi$ where
$c_{\rm eff}$ counts the effective degrees of the system. For our case,
$c_{\rm eff}=2$ counting both the spin up and down degrees of
freedom.
The fermion numbers and flux numbers from the gravity side are
respectively given by
\be
Q'_e=\sigma_F Q_e = \rho_e \, \gamma \, L^2 \qquad {\rm and}\qquad
Q'_m=\sigma_F Q_m = \rho_m \, \gamma \, L^2\,.
\ee
Here $\sigma_F$ corresponding to dialing $e_G$ is the conversion factor
between the gravity charges and particle numbers in the CFT side.
We use $\sigma_F=2/\sqrt{3}$ for the match to the weakly coupled
conformal fermions.

Setting the mass density $m =(k_0)^3$, one finds $\rho_e^2+ \rho^2_m= 3 k_0^4$
from the saturation of the cosmic censorship inequality in (\ref{inequality})
at zero  temperature. This then leads to
\bea
 E_g&=& k^3_0\, \gamma \, L^2\nonumber\\
 \sqrt{Q_e^2 + Q_m^2} &=& {3\over 2} \,  k^2_0\, \gamma \, L^2\,.
\label{q1}
\eea
We  note that fermi momentum for the composite fermion is
given by
\be
Q_e= {3\over 2} \,  k^2_F\, \gamma \, L^2\,
\label{qe}
\ee
and $Q_m = 2p Q_e$
since $2p$ flux quantum is attached to
each electron. Comparing (\ref{q1}) with (\ref{qe}), we get
$k_0= k_{\rm F} \, (1+ 4p^2)^{1/4}$. Consequently, the ground state energy of the
composite fermion metal is found to be
\be
E_G = {2\over 3}\, Q_e\, k_{\rm F} (1+ 4p^2)^{3\over 4}
= E_0 (1+ 3p^2 +\cdots)\,.
\label{groundenergy}
\ee
This agrees perfectly with the scaling of the leading Hartree-Fock approximation (\ref{HFgs}).
In comparing the two results, we stress that, in the composite fermion, details of the constituent
fermion are largely irrelevant. In particular, whether the constituent fermion is Galilean invariant (which has a fixed band mass $m$) or Lorentz invariant ought not to matter.

We emphasize that the above agreement is a direct confirmation of the composite Dirac fermion picture.
To substantiate this further, consider an alternative
interpretation
that the magnetic field of the
dyon black hole drives the system to a quantum Hall state. In this case, we may estimate the ground-state energy using the substitution rule in section 2 that replaces fermion's band mass $2 m_*$ to
$E_{\rm HF} / c^2$.
Recalling the Fermi energy $E_{\rm F} =
\hbar^2 k_{\rm F}^2/ 2
m_*$ for fermions with band mass $m$, we find
\bea
E_{\rm HF} = \sqrt{N} \hbar k_{\rm F} c \ \sqrt{ 1 + 3 p^2}
\eea
for Dirac composite fermions. This does not agree with the dyon black hole energy even for $p=0$.
This mismatch simply implies that $m$ in the gravity side
cannot be interpreted as a band mass $m_*$.

Further support for the conformally invariant composite Dirac fermion metal comes from the finite temperature corrections. From the gravity side, it is straightforward to
identify the temperature dependence of the energy by considering finite-temperature dyon black hole:
\be
E= E_G \,\left[\, 1 +
{1\over 2} \left( {\pi T\over k_{\rm F} (1+ 4p^2)^{1\over 4}}
\right)^2 + \cdots \right]\,.
\ee
This shows that the specific heat of the system is indeed linear in $T$, thus exhibiting Fermi liquid behavior \footnote{The linear $T$ behavior of the specific heat is actually because the ground-state entropy of the dyon black hole is nonzero and large, in stark difference from Fermi liquid. At present, this 'entropy enigma' is an open problem that underlies all holographic proposals to condensed-matter systems
including ours. Dilatonic anti-de Sitter black holes have vanishing entropy and hence may provide a more realistic setup for a holographic dual of the composite fermion metal. On the other hand, some other physical observables seem to match between two sides of the proposed holography. Static structure factor at zero and low temperatures is one such example \cite{bakreyFS}. }. This expression also suggests that the combination $E_F= k_F (1+ 4p^2)^{1/4}$ plays the role of Fermi energy for the composite fermion metals, both at zero and finite temperature.

\section{Holography of Cyclotron Motion}
A crucial and non-perturbative feature distinguishing composite fermions from constitute electrons
is that composite fermions experience an effective magnetic field $B^*$ which is drastically different
from the external magnetic field $B$ the constituent electrons experience. The effective magnetic field
is a direct consequence the formation of composite fermions, so its observation would be tantamount to
an observation of the composite fermions themselves.

The distinction becomes most dramatic for $\nu = 1/2$ half-filling. At this filling, the external magnetic field $B = 2 \rho \Phi_0$ is enormous, causing the constituent electrons to undergo cyclotron motion, while the effective magnetic field $B^*$ vanishes identically, letting the composite fermions drift freely. Therefore, a picture based on the mean-field approximation suggests that composite fermions form a Fermi surface.

Halperin, Lee and Read \cite{HLR} argued that many features of the Fermi surface persists even after fluctuations beyond the mean-field approximation are taken account of. Thus, if $\nu$ is slightly above $1/2$, the composite fermions would experience a very weak effective magnetic field $B^*$ and circulate around a classical cyclotron orbit of effective radius $R^* = \hbar k_{\rm F} / e B^*$, where
the Fermi momentum is given by $k_{\rm F} = \sqrt{2 \pi \, \rho }$ with the number density
$\rho\equiv Q_e/L^2$. On the other hand,
constituent electron will see the external magnetic field $B$ and execute cyclotron orbit of radius $R = \hbar k_{\rm F} / e B$. Having $\nu$ slightly away from $1/2p$,
we have that $R^* \gg R$. Notice that, in this line of reasoning, we have taken electric charge of the composite fermion the same as electric charge of the constituent electron.

We shall now introduce a probe fermion in the bulk and study its classical cyclotron motion. In the dual system, this amounts to creating a fermionic quasi-particle excitation by a gauge-invariant operator of appropriate scaling dimension. By construction, the operator is neutral with respect to the statistical
gauge potential $a_m$ but still can couple to the external gauge potential $A_m$. Thus, by tuning the external magnetic field, we are probing cyclotron motion of the fermionic quasi-particle. Experimentally,
these are already measured with precision \cite{cyclotron}.

According to the AdS/CFT correspondence, a massive field in the bulk corresponds to a gauge-invariant
operator of scaling dimension in the boundary conformal theory. Here, we have a novel situation that
the composite operators involve not only constituent electrons and holes but also magnetic fluxes. The
fermion number of electrons and holes is the same as the charges of (2+1)-dimensional gauge charges.
The quantum number of magnetic fluxes is the same as the (2+1)-dimensional magnetic field. Therefore,
by the operator-field map in the AdS/CFT correspondence, we identify the gauge invariant operator
${\cal O}^\dagger$ creating a composite fermion in (2+1) dimensions with the fermion field $\Psi$
coupled minimally to both electric and magnetic gauge potentials in the Einstein-Maxwell system.
Thus, the action is given by
\bea
I = \int \rmd^4 x \sqrt{-g} {1 \over 16 \pi G_4}
\Big[ \overline{\Psi} \, e_a^m\, \gamma^a\,
 D_m \Psi - m_\Delta
\overline{\Psi} \Psi \Big]
\eea
where the covariant derivative is defined by
\bea
D_m \Psi(x) = \partial_m \Psi + {1 \over 8} {\omega_m}^{ab}[\gamma_a, \gamma_b] \Psi
- i (q A_m + \tilde{q} \widetilde{A}_m) \ \Psi
\eea
As said, the fermion field couples minimally to
 both gauge potential $A_m$ and dual
gauge potential $\widetilde{A}_m$, where the two are related each other
by the Poincar\'e duality
\bea
\tilde{F}_{mn}={1 \over 2} {\epsilon_{mn}}^{pq} F_{pq},
\eea
together with
\bea
F_{mn} = \partial_m A_n - \partial_n A_m \qquad \mbox{and} \qquad
\tilde{F}_{mn}= \partial_m \tilde{A}_n - \partial_n \tilde{A}_m\,.
\eea
To avoid double counting, we shall thus choose gauge potential to couple to the electric charge
of the fermion field $\Psi$ and dual gauge potential to couple to the magnetic charge only.
\begin{figure}[ht!]
\vskip1cm
\centering
\includegraphics[scale=0.75]{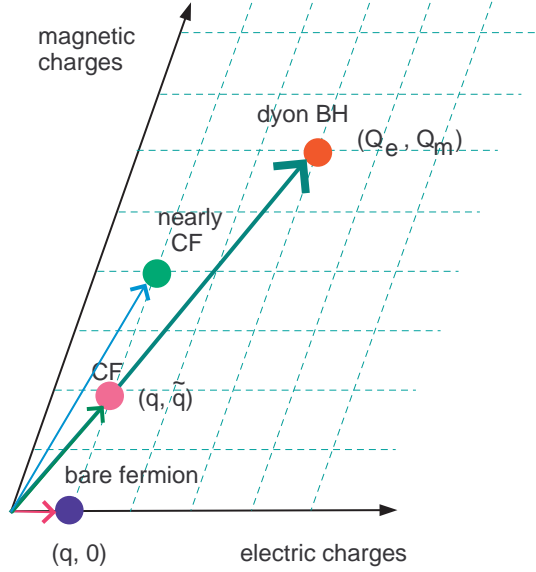}
\caption{\small \sl
Charge lattice for states in (3+1) dimensions and operators in (2+1) dimensions. For AdS$_4$ states, horizontal and vertical axis represent electric and magnetic charges $(q, \tilde{q})$. They corresponds to CFT$_3$ operators carrying fermion number $q$ and statistical flux quanta $\tilde{q}$. Dyon black hole, composite fermion and constituent fermions are marked by red, purple and blue circles, respectively. A probe fermion state (green circle) whose charge vector is slightly disoriented from charge vector of dyon black hole feels at mean field approximation a weak effective magnetic field. }
\label{}
\vskip1cm
\end{figure}
The electric and magnetic charges $(q, \tilde{q})$ of the fermion field $\Psi$ is identified with the
fermion number and magnetic flux number $\Phi/\Phi_0$ of the gauge-invariant composite fermion creating
operator ${\cal O}^\dagger$ in (2+1)-dimensions. Also, the mass $m_\Delta$ of the fermion field $\Psi$ is
identified with the scaling dimension $\Delta$ of the operator ${\cal O}^\dagger$:
\bea
m_\Delta = \Delta - {3 \over 2}.
\eea

Under the $SL(2, \mathbb{Z})$ S-duality, the bulk fermion field transforms the same way as the black hole
itself. Thus,
\bea
&& S: \hskip1.2cm q \rightarrow \tilde{q} \hskip2.3cm \mbox{and} \qquad
\tilde{q} \rightarrow - q \nonumber \\
&& T: \hskip1.1cm q \rightarrow q + 2p \ \tilde{q} \hskip1.1cm
\mbox{and} \qquad \tilde{q} \rightarrow \ \tilde{q}.
\eea

We shall solve the Dirac equation in the background of the dyon black hole. First, consider
a bulk fermion field $\Psi_{q, 0}$ carrying electric charge only. This corresponds in the composite fermion metal to a gauge
invariant operator creating a constituent electron and hole. We treat them in the probe approximation, thus ignore back-reaction to the change of the electron density $\rho_e$ and hence filling factor $\nu$. In the
composite fermion metal, in the mean-field picture, we expect the electron to undergo a cyclotron motion.
We shall find that the bulk fermion also undergoes a motion whose projection to $(x,y)$-plane is precisely
the cyclotron motion.
For this purely electric coupling,
the explicit form of the bulk Dirac equation  is given by
\be
\Big[\gamma^z \big(\partial_z -{f'\over 4f} -{1\over 2z}\big)
+ {\gamma^t \over \sqrt{f}} \big(\partial_t - iq A_t\big)-{m_\Delta\over z}
+\gamma^x  \big(\partial_x - iq A_x\big)
+\gamma^y  \big(\partial_y - iq A_y\big)
\Big]\Psi=0\,.
\label{probe}
\ee
If $A_i=0\,\,\, (i=1,2)$ with $A_t$ such that
$F_{tz}=\rho_e/4$, we have shown that Fermi sea filled up to the Fermi
level is probed as discussed in Ref.~\cite{bakreyFS}. The operator
$\gamma^i \,\partial_i$ is diagonalized with the usual momentum
eigenstate in the $(x,y)$ directions. This is quite consistent with
the known spectra of the boundary Dirac fermion. If one now
introduces a small bulk $B$ field, the operator $\gamma^i\, D_i$ is still
completely independent of the bulk coordinate $z$ and simply corresponds to
the flat transverse space operator  minimally coupled to the
constant magnetic field $B$. It is diagonalized by the eigenfunctions
representing the cyclotron motion in the semiclassical limit.

To be explicit, we introduce a normalization of coupling
$q$ such that $qA_i = -eB \epsilon_{ij}x_j$ in the symmetric
gauge. This corresponds to the magnetic field
$q \epsilon_{ij}\partial_i A_j/e= B$ in the boundary
side. The relevant part of $4 \times 4$ Dirac matrices are given by
$\gamma^i = \sigma_1 \otimes \sigma_i$ so that
the operator $i\gamma^i D_i$ takes a form
$ \sigma_1 \otimes\, i \sigma_i D_i$.   We further introduce
a set of creation and annihilation operators by
\bea
\alpha^\dagger = i \sqrt{2\over eB} \Big(
\partial_{{w}} -{eB\over 4} \bar{w}
\Big)\,, \ \ \ \  \alpha = i \sqrt{2\over eB} \Big(
\partial_{\bar{w}} +{eB\over 4} w
\Big)\,,
\eea
and
\bea
\beta^\dagger = -i \sqrt{2\over eB} \Big(
\partial_{\bar{w}} -{eB\over 4} {w}
\Big)\,, \ \ \ \  \beta =- i \sqrt{2\over eB} \Big(
\partial_{{w}} +{eB\over 4} \bar{w}
\Big)\,,
\eea
where $w$ is the complex coordinate $x+iy$.
The normalized eigenfunction diagonalizing
$\alpha^\dagger\alpha$ and
$\beta^\dagger\beta$ is given by
\be
\phi_n^l(x)=\sqrt{ (eB)^{l+1}\,\,n!\over 2^{l+1}\,
\pi\,\,(n+l)!}\, w^l\, L^l_n
\Big({eB\over 2} |w|^2\Big) e^{-{1\over 4}eB \ |w|^2}\,
\ee
where $L_n^l(x)\,\,(l \ge -n)$
is the generalized Laguerre polynomial. The creation and annihilation
operators acting on the above eigenfunction are as follows:
\bea
&&
\alpha^\dagger\phi_n^l =i\sqrt{n+1}\,\,\phi^{l-1}_{n+1}\,,
\ \ \ \ \ \ \ \
\alpha\, \phi_n^l =-i\sqrt{n}\,\,\phi^{l+1}_{n-1}\,, \nonumber\\
&&\beta^\dagger\phi_n^l =i\sqrt{n+l+1}\,\,\phi^{l+1}_{n}\,, \ \ \
\beta\, \phi_n^l =-i\sqrt{n+l}\,\,\phi^{l-1}_{n}\,.
\eea
We see that the state $\phi_n^l$ is annihilated to zero by
$\alpha\,\,/\,\,\beta$ if $n=0\,\,/\,\, l=-n$.
Orbital angular momentum is defined by ${L}=
-\alpha^\dagger\alpha+\beta^\dagger\beta$ and $l$ is the
corresponding quantum number. Now noting
\bea
i\sigma_i D_i =\sqrt{2 e B}\,\,\left[
\begin{array}{cc}
0 & \alpha^\dagger\\
\alpha & 0
\end{array}\right]\ ,
\eea
we find wave functions
\bea
|\phi^{\pm}_{nj}\rangle=
{1\over \sqrt{2}}\,\,\left[
\begin{array}{c}
 \phi^{j-\mbox{\small ${1\over 2}$}}_{n+1}\\
\phi^{j+\mbox{\small ${1\over 2}$}}_{n}
\end{array}\right]
\eea
diagonalizes $i\sigma_i D_i$ with eigenvalues
$\pm \sqrt{eB}\,\sqrt{2(n+1)}$.
 The quantum number
$j$ is the eigenvalue of the total angular momentum
\be
J= {L} + S= {L} + {1\over 2}\,\sigma_3\,,
\ee
and takes half integer
values $j\ge -n-\mbox{\small ${1\over 2}$}$. Introducing further
the basis $\sigma_1 |\pm\rangle =\pm |\pm\rangle$, the eigenfunction
of the operator $i\gamma^iD_i$ is given by
\be
 i\gamma^iD_i\,\, |\pm\rangle \otimes |\phi^{\pm}_{nj}\rangle=
(\pm)(\pm)\sqrt{eB}\sqrt{2(n+1)}\,\,
|\pm\rangle \otimes |\phi^{\pm}_{nj}\rangle\,.
\ee
Therefore we conclude that an additional bulk magnetic
field (without the back-reaction included) subjects the fermi-sea
electrons to cyclotron motion in their semiclassical picture. Their
radius is $R=\hbar k/eB$ where $\hbar k$ is the transverse momentum
in the original momentum basis.

We now turn to the composite fermion in which a constituent Dirac fermion is attached
with $2p$ flux quanta, viz. $\tilde{q}= 2p \, q$.
At general filling fraction $\nu$,
we have
\be
F= {\rho_e\over 4}\, \rmd t \wedge  \rmd z +  {\rho_e\over 4\, \nu}\,\, \rmd x\wedge \rmd y
\ee
in the unit we are using. The dual field strength is then
\be
\tilde{F}= {\rho_e \over 4\nu}\,\, \rmd t \wedge  \rmd z -{\rho_e\over 4}\, \rmd x\wedge \rmd y \ .
\ee
In this case, the bulk probe fermion field carrying the same quantum numbers is a gauge-invariant
description of single composite fermion as a quasi-particle excitation in the composite fermion metal. In this sense, we may regard the bulk fermion field is a dyon field except it carries the statistics of the
fermion. But then, we note that the bulk fermion field experiences effectively
the bulk gauge field
\be
F_{\rm eff}
=F + {\tilde{q}\over q}\, \tilde{F}
={\rho_e\over 4} \,
\Big(
1+{2p\over \nu}\,\Big)\, \rmd t\wedge \rmd z + {\rho_e\over 4}
 \,\Big({1\over \nu} - 2p\Big)
\,\,\rmd x \wedge \rmd y\,.
\label{magneticfield}
\ee

We are interested in the metallic point $\nu=1/2p$. From (\ref{magneticfield}), we see that the effective magnetic field experienced by the bulk fermion field $\Psi_{q,p}$ vanishes completely, viz.
\be
F_{\rm eff}= {\rho_e\over 4} \,
\big(
1+4p^2\,\big)\, \rmd t\wedge \rmd z\,.
\ee
Accordingly, in solving the Dirac equation corresponding to gauge invariant operator creating a composite fermion,
The $q\, A_m$ in 
(\ref{probe}) is to be replaced by $q\, A_m^{\rm eff}$.
Thus, without much work, we see that the bulk fermion field $\Psi_{q,p}$ does not experience any magnetic component of the gauge potential, viz. spatial part of the wave function is simply plane-waves in the $x-y$-directions. This fits consistently with our picture that the composite fermion system is a metal with {\sl no} cyclotron motion. It probes only the Fermi sea with the Fermi level scaled accordingly, which is not influenced by any (residual) magnetic field.

Consider now the more general filling fraction $\nu=k/(2pk+1)$. The bulk fermion field $\Psi_{q,p}$ experiences the reduced magnetic field
\be
B_* = B - 2p \, \rho \, \Phi_0
=\Big(
{1\over\nu}-2p
\Big) \rho\, \Phi_0 = {1\over k } \rho\, \Phi_0 \, .
\ee
This effective magnetic field can be made arbitrarily weak by letting the integer $k$ large.
Therefore, we conclude that the composite fermions is semiclassically experiencing the cyclotron motion
with radius:
\be
R_* = \hbar k_{\rm F} / e B_*
\eea
at the Fermi level.

\section{Discussions}
In this work, we proposed that gravity in the background of dyon black hole in the four-dimensional anti-de Sitter spacetime corresponds to quantum critical system of composite Dirac fermions at filling fractions $1/2p$. We have shown that the system exhibits thermodynamic behavior of ideal Fermi liquid by
relating the dyon black hole via the $\Gamma_0(2)$ orbit of S-duality map to the purely electric black hole and by recalling our earlier result that the latter black hole describes conformally invariant Fermi liquid. We found further supporting evidences by matching the ground state energy, linear specific heat at low temperature and the cyclotron motion of various (composite) fermions in a weak magnetic field background.
Finer details of thermodynamics and transport properties are somewhat different from the analysis based on
fermionic Chern-Simons theory and experiments but we believe they are attributed to the fact that the composite fermions under consideration is relativistic Dirac fermions rather than non-relativistic Pauli fermions.

We conclude with discussions of several issues that deserve further study.
\begin{list}{$\bullet$}{}

\item
The nonperturbative wave function for the composite fermion is proposed to take the form
\be
\Phi_{2p}= \prod_{a < b}^N(z_a-z_b)^{2p} \ \Phi
\ee
where $z_a = x_a + i y_a$ denotes the position of $a$-th
massless fermion as a complex number and $\Phi$
is the Slater-determinant wave function for the
noninteracting Dirac fermion system. It would be interesting
to see if the energy evaluated with this wave function
agrees with our gravity prediction in (\ref{groundenergy}).
The above type of  wave function is not exact but turned
out extremely accurate approximation of the real
ground-state wave function for Galilean invariant systems.

\item It would be helpful to understand how to turn on
residual magnetic field in the bulk beyond the probe approximation
we have used. One suggestion we put forward is to introduce
bulk magnetic source at an appropriate location {\sl outside}
the horizon.  This then has two-fold effects on the system.
First, the bulk location of the magnetic source will introduce
additional scale which may correspond to the cyclotron
frequency. Second, the shape of the black hole inside
this source location can be approximately intact representing
$(1,2p)$ composite fermion system. Detailed understanding of such
configuration would reveal many interesting facets of the proposed
composite fermion metals. Related, to substantiate our proposal further,
it is desirable to study other physical observables in the composite
fermion metal system.

\item Thermodynamic behavior of the dyon black hole and, via holography, of the composite Dirac fermion metal parallels well with the ideal Fermi liquid. However, all the proposed holographic duals of charged black hole (including ours) have some outstanding pitfalls. First, the ground-state entropy is not controllable -- either it is too large (in unit of the Fermi energy) for all non-dilatonic charged black holes or absolutely zero for all dilatonic charged black holes. Second, in string theory realization of these black holes, microstates of holographic dual contain not only massless fermions but also massless bosons. It is unclear why the black hole does not exhibit Bose-Einstein condensation of the massless bosons, expected at weak coupling regime of the holographic dual. Our proposal shares with all other proposed holography of fermi system for being built upon these pitfalls. Understanding and resolving these 'entropy and boson enigmas' are extremely interesting and rewarding issues left for future study.

\item In high mobility semiconductors, it was observed experimentally that the Fermi velocity vanishes \cite{Fvelocity}, implying that the effective mass of the composite fermions diverges at zero temperature. In our consideration, the constituent fermions are Dirac fermions and differs from the situation of high mobility semiconductors. Nevertheless, there are reasons to believe that divergent effective mass is a feature largely insensitive to detailed nature of constituent fermions. Holographic understanding the fermi velocity, equivalently, effective mass of the composite Dirac fermion would thus be very interesting. This will also shed light how dyon black hole thermodynamics may depart from that of ideal Fermi liquid.
    In this regard, it should be borne in mind that quantum and thermal limits in general do not commute. By correspondence principle, a quantum degenerate Fermi gas approaches a Maxwell-Boltzmann gas at high temperature. The zero temperature limit of the Maxwell-Boltzmann gas does not go back to the quantum degenerate Fermi gas.

\item There has been suggestions that true quasi-particles in the composite fermion metals are dipoles in that excitations are effectively neutral but still carries electric dipole moments \cite{dipole}. It would be interesting to construct holographic dual of these dipoles in the dyon black hole. Interpreting the background magnetic field on $\mathbb{R}^2$ as a source of noncommutative space, it was previously noted that the open Wilson lines \cite{OWL} are essentially Mott excitons and behave much similar to such dipoles \cite{reylecture}. It would be very interesting to make the connection more concrete and complete by implementing the effect of dyon black hole background.

\item It would be interesting to understand if the dyon black hole is capable of describing an exotic class of quantum Hall states which might arise when certain aspects of BCS pairing is combined with Laughlin-type ordering. The best known exception of this sort is the $\nu = 5/2$ state. In so far as quantum numbers are concerned, we can straightforwardly construct the corresponding dyon black hole from the S-duality map. For example, starting from a purely electric black hole which describes a Fermi liquid of constituent fermions and making $ST^{2q}ST^{2p}S$ transformation belonging to the orbit of $\Gamma(2)$, we can construct the filling fraction $\nu = (4 pq + 1 )/ 2 p$. Notable feature of the $\nu = 5/2$ state is that ground-state is incompressible and quasiparticles exhibit nonabelian braiding statistics. The ground-state is described by a Pfaffian state by Moore and Read \cite{MR}. Such novel statistics may give rise to a source of rich topological quantum entanglement and, for this reason, it has attracted much attention for the prospect of topological quantum computations. It is yet unclear how such nonabelian structure is embedded into the hierarchy of the dyon black hole. Unveiling nonabelian statistics and novel topological quantum entanglement from black hole physics would be extremely fascinating. With rapid progress on the $\nu = 5/2$ Pfaffian state and the topological quantum computations, we find it particularly exciting that string theory has much to learn and gain insight on black hole physics from condensed matter physics.

\item
The approach we have taken in this paper is largely phenomenological. To identify the $(2+1)$-dimensional conformal field theory of composite fermion metals, it is imminent and utmost important to embed the dyon black hole or suitable variant of it in precise and controlled ways to string theory on AdS$_4 \times X_6$ for a suitable six-dimensional compact manifold $X_6$ or to M-theory on AdS$_4 \times X_7$ for a suitable seven-dimensional compact manifold $X_7$. A minimal requirement for these internal manifolds $X_6$, $X_7$ is that the Kaluza-Klein spectrum should contain massless fields that parametrize $P, T$ violating coupling of Maxwell fields on AdS$_4$ spacetime. Moreover, the choice of $X_6$ or $X_7$ permits as a dual to the compactification an explicit Lagrangian formulation of $T$-violating three-dimensional conformal field theory that is continuously deformable to the fermionic Chern-Simons theory we discussed in section 2.
Recent work in \cite{gauntlett} may shed some hints on this issue.

\end{list}

We are currently working on these issues and will report our progress elsewhere.

\section*{Acknowledgement}
We thank P.W. Anderson, G. Baskaran, E. Fradkin, J.K. Jain, J. Maldacena, N. Nagaosa and K. Yang for many useful discussions and correspondences.
This work was supported by the National Research Foundation of Korea Grants 2005-0049409 (CQUEST), KRF 2005-084-C00003, 2009-008-0372, KOSEF R01-2008-000-10656-0, EU-FP Marie Curie Research \&
Training Networks HPRN-CT-2006-035863 (K209090-00001-09B1300-00110) and the U.S. Department of
Energy Grant DE-FG02-90ER40542.

\end{document}